\documentclass[twocolumn,showpacs,preprintnumbers,superscriptaddress,amsmath,amssymb,prl]{revtex4}

\usepackage{amssymb}
\usepackage{stmaryrd}
\usepackage{amsmath}
\usepackage{txfonts}
\usepackage{graphicx}
\usepackage{CJK}

\usepackage{bm}
\usepackage{ulem}

\begin{document}
\begin{CJK*}{GB}{SongMT}
\CJKfamily{gbsn}

\title{Near-complete polarization Bell-state analysis based on symmetry-broken scheme with linear optics}

\author{Ling-Jun Kong}
\affiliation{School of Physics and Key Laboratory of Weak-Light Nonlinear Photonics, Nankai University, Tianjin 300071, China}
\author{Yongnan Li}
\affiliation{School of Physics and Key Laboratory of Weak-Light Nonlinear Photonics, Nankai University, Tianjin 300071, China}
\author{Yu Si}
\affiliation{School of Physics and Key Laboratory of Weak-Light Nonlinear Photonics, Nankai University, Tianjin 300071, China}
\author{Rui Liu}
\affiliation{School of Physics and Key Laboratory of Weak-Light Nonlinear Photonics, Nankai University, Tianjin 300071, China}
\author{Chenghou Tu}
\affiliation{School of Physics and Key Laboratory of Weak-Light Nonlinear Photonics, Nankai University, Tianjin 300071, China}
\author{Hui-Tian Wang}
\email{htwang@nju.edu.cn/htwang@nankai.edu.cn}
\affiliation{School of Physics and Key Laboratory of Weak-Light Nonlinear Photonics, Nankai University, Tianjin 300071, China}
\affiliation{National Laboratory of Solid State Microstructures, Nanjing University, Nanjing 210093, China}
\affiliation{Collaborative Innovation Center of Advanced Microstructures, Nanjing University, Nanjing 210093, China}

\date{August 19, 2014}

\begin{abstract}
\noindent
Bell-state analysis is a considerable challenge and an essential requirement for reliable implementation of quantum communication proposals. An open question is the one for the maximal fraction of successful Bell measurements. It has been pointed out that no scheme using only linear elements can implement a Bell state analyzer. Some effort has paid attention to the complete polarization-entangled Bell-state analysis using linear optics, with the aid of auxiliary means. Here we present a symmetry-broken scheme with linear optics only, without any aid of other auxiliary means, for discriminating polarization-entangled Bell states. Although our scheme is unable of realizing complete Bell-state measurement for less photon-pairs situation, it can deterministically identify four Bell states with success probabilities beyond 99.2\% provided that photon-pairs are not less than 8. Our scheme as a significant breakthrough is simpler and feasible with respect to the current technology for the near-complete Bell-state analysis. Symmetry breaking is indispensable in our scheme as in other physical systems.
\end{abstract}


\maketitle
\end{CJK*}


\noindent Studies on correlation and entanglement between quantum systems were initiated by Einstein, Podolsky, and Rosen (EPR) to inquire the completeness of quantum mechanics \cite{R01}. For a long time, entangled states have been regarded as tools for studying fundamental issues in quantum mechanics \cite{R02, R03}. Entangled states have formed the cornerstone of the newly emerging field of quantum information. Numerous potential uses of quantum correlations for quantum computation and for quantum information have been discovered \cite{R04}. Entangled states have largely improved the methods of manipulating and transforming information. Entangled photon systems can function as quantum channels in some typical long-distance quantum communication proposals, such as quantum key distribution \cite{R05}, quantum dense coding \cite{R06}, quantum teleportation \cite{R07, R08,R09}, entanglement swapping \cite{R07, R10, R11}, and quantum secret sharing \cite{R12}. Because Bell states are the key ingredients in all these quantum communication proposals, Bell states have to be prepared and measured. The problem of preparing Bell states has been solved by using parametric down-conversion in a nonlinear crystal \cite{R13, R14, R15}. Particular Bell states can be prepared from any maximally entangled pair through simple local unitary transformations. Whether or not it is possible to perform a complete Bell-state analysis for distinguishing between the four maximally entangled Bell states is a question that needs to be addressed.

The simplest Bell states are the polarization-entangled Bell states that can be achieved through the nonlinear interactions involved in spontaneous parametric down-conversion \cite{R13, R14, R15}. For the polarization Bell-state analysis, many theoretical proposals and experiments have been done. In 1996, Michler \textit{et al}. \cite{R16} realized experimentally the Bell-state analysis scheme, in which some linear optical elements were utilized and the probability of success was 50\%. These results indicate that only two Bell states can be distinguished from the four Bell states. Much attention has been devoted to identifying the four Bell states. For instance, using the additional degrees of freedom \cite{R17}, which refer to the hyperentangled states \cite{R18, R19}, can achieve the four Bell-states analysis. Other similar Bell-state analysis schemes have been proposed \cite{R20,R21,R22}. Moreover, several experiments on hyperentanglement have been performed \cite{R23,R24}. Based on the nonlinear interaction, a Bell-state measurement also has been realized \cite{R25}. Some effort has been exerted to identify the polarization-entangled Bell states using linear optics, with the aid of auxiliary photons, feedback technique, and/or nonlinear sign-shift gates \cite{R26,R27,R28}. It has been pointed out that no scheme using only linear elements can implement a Bell state analyzer \cite{R29,R30}. Although a symmetric scheme using linear optical elements only was proposed to attempt to discriminate all four Bell states recently\cite{R31}, it was subsequently proven to be futile \cite{R32}. However, we still have a question that whether or not the four Bell states cannot indeed be identified using linear optics only, without any aid of other auxiliary means. Strict symmetry may be boring, but symmetry breaking has acquired special significance in physics and is in this sense what ``creates the phenomenon". The Bell-state analysis is a considerable challenge and an essential requirement for reliable implementation of quantum communication proposals. Therefore, we introduce the symmetry breaking, which is expected to identify the polarization Bell states completely.

In this article, we present a symmetry-broken scheme using linear optics only, for near completely identifying all the four polarization Bell states of entangled qubits. We employ linear optics only, including 50/50 beam splitters, polarization beam splitters, and half-wave plates, without any aid of other auxiliary means (such as auxiliary photons, feedback technique, and nonlinear sign-shift gates). Although our scheme is not capable of implementing the completely deterministic Bell-state measurement for less photon-pairs situation, it can deterministically identify the four Bell states with success probabilities beyond 99.2\% when photon-pairs are not less than 8. Since such a level of photon-pairs is much lower than the practically used level in the current experiments \cite{R16,R23,R24,R25,R27,R28}, our scheme should be very useful and beneficial for the polarization Bell-state analysis under the present experimental condition. Our work should be a crucial breakthrough in the entangled Bell-state analysis and of great significance in practical quantum applications. Our scheme is simpler and feasible with respect to the current technology for the Bell-state analysis. Symmetry breaking has an indispensable function in our scheme as in other physical systems, and may serve as an important reference for characterizing other types of quantum states.

\noindent \textbf{Principle.} For the polarization degree of freedom, the maximally entangled Bell states of two photons are
\begin{subequations}
\begin{align}\label{01}
| \psi ^{-}\rangle  = & \frac{1}{\sqrt{2}} (|H \rangle_{a'} |V \rangle_{b'} - |V \rangle_{a'} |H \rangle_{b'} ), \\
| \psi ^{+}\rangle  = & \frac{1}{\sqrt{2}} (|H \rangle_{a'} |V \rangle_{b'} + |V \rangle_{a'} |H \rangle_{b'} ), \\
| \phi ^{-}\rangle  = & \frac{1}{\sqrt{2}} (|H \rangle_{a'} |H \rangle_{b'} - |V \rangle_{a'} |V \rangle_{b'} ), \\
| \phi ^{+}\rangle  = & \frac{1}{\sqrt{2}} (|H \rangle_{a'} |H \rangle_{b'} + |V \rangle_{a'} |V \rangle_{b'} ),
\end{align}
\end{subequations}

\noindent where $|H \rangle$ and $|V \rangle$ denote horizontal and vertical polarization, respectively. Among the above four Bell states, only $| \psi ^{-}\rangle$ is an antisymmetric singlet state, while the other threes are symmetric triplet states because they follow the permutation symmetry. When photons meet at a 50/50 beam splitter (BS) that can perform the unitary $U(2)$ operation, the input modes $|X \rangle_{a'}$ and $|X \rangle_{b'}$ will be converted into $|X \rangle_{a'}\Rightarrow {1 \over \sqrt{2}} (|X \rangle_{a}+|X \rangle_{b})$ and $|X \rangle_{b'}\Rightarrow {1 \over \sqrt{2}} (|X \rangle_{a}-|X \rangle_{b})$ (where $X = H$ or $V$), respectively. The Hong-Ou-Mandel interference \cite{R31} will then occur. The overall bosonic symmetry of the two-photon state requires that photons in the antisymmetric singlet state exit in different output ports, whereas photons in the symmetric triplet state end up in the same output port \cite{R33,R34} which is the so-called photon bunching effect. Based on this effect, $| \psi ^{-}\rangle$ can be separated from the others. Furthermore, with a polarization beam splitter (PBS), $| \psi ^{+}\rangle$ can be distinguished from $| \phi ^{\pm}\rangle$. However, $| \phi ^{+}\rangle$ and $| \phi ^{-}\rangle$ cannot be still distinguished from each other.

Transforming one Bell state into another is easy. For instance, if the initial state is $| \psi ^{+}\rangle$, (i) with the aid of the polarization exchange ($| H \rangle \Rightarrow | V \rangle $ and  $| V \rangle \Rightarrow | H \rangle $) which is realized by a half-wave plate (HWP), $| \psi ^{+}\rangle$ can be changed into $| \phi^{+}\rangle$; (ii) with the aid of the polarization-dependent phase shift generated by a quarter-wave plate \cite{R06}, $| \psi ^{+}\rangle$ can be transformed to $| \psi ^{-}\rangle$; (iii) with the aid of both polarization exchange and polarization phase shift, $| \psi ^{+}\rangle$ can become $| \phi^{-} \rangle $. As a result, by using only the polarization exchange, we can transform the two Bell states $| \psi ^{+}\rangle$ and $| \psi ^{-}\rangle$ into $| \phi^{+}\rangle$ and $| \phi^{-}\rangle$, respectively. As discussed above, $| \psi ^{+}\rangle$ and $| \psi ^{-}\rangle$ can be first distinguished. Then, if $| \phi ^{\pm}\rangle$ can be transformed into $| \psi ^{\pm}\rangle$, based on the Hong-Ou-Mandel interference \cite{R31} and the photon bunching effect, $| \phi^{+}\rangle$ and $| \phi^{-}\rangle$ become also distinguishable from each other.
\begin{figure}
\centering{\includegraphics[width=8.5cm]{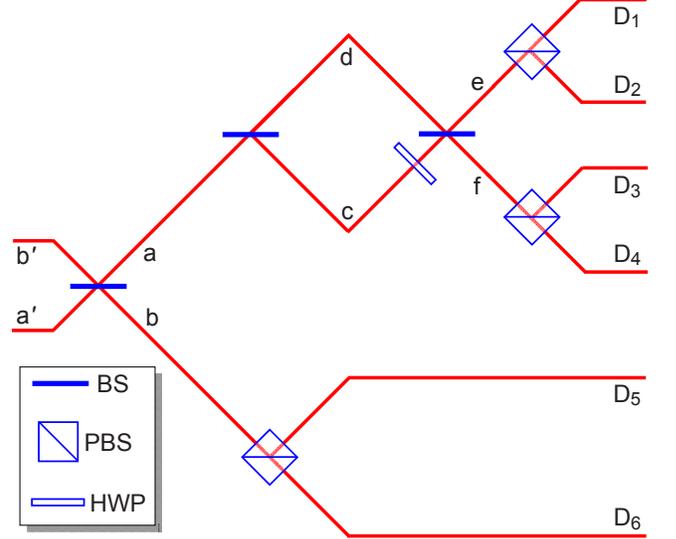}\\
  \caption{Symmetry-broken scheme with linear optics for identifying all four polarization Bell states. BS---50/50 beam splitter, PBS---polarization beam splitter, HWP---half-wave plate.}}
\end{figure}

As shown in Fig.~1, our scheme for the near-complete polarization Bell-state analysis breaks indeed the symmetry in the geometric arrangement and the types (functions) of the linear optical elements. Two photons ($\mathbf{a}'$ and $\mathbf{b}'$) reach a BS from its opposite sides and interfere, and then evolve into two other photons ($\mathbf{a}$ and $\mathbf{b}$). Thus, the four Bell states described in Eq.~(1) become
\begin{subequations}
\begin{align}\label{02}
| \psi ^{-}\rangle \Rightarrow & \frac{1}{\sqrt{2}} (|H \rangle_b |V \rangle_a - |V \rangle_b |H \rangle_a ), \\
| \psi ^{+}\rangle  \Rightarrow & \frac{1}{\sqrt{2}} (|H \rangle_a |V \rangle_a - |H \rangle_b |V \rangle_b ), \\
| \phi ^{-}\rangle  \Rightarrow & \frac{1}{2\sqrt{2}} (|H \rangle_a |H \rangle_a - |V \rangle_a |V \rangle_a -|H \rangle_b |H \rangle_b + |V \rangle_b |V \rangle_b  ), \\
| \phi ^{+}\rangle  \Rightarrow & \frac{1}{2\sqrt{2}} (|H \rangle_a |H \rangle_a + |V \rangle_a |V \rangle_a -|H \rangle_b |H \rangle_b - |V \rangle_b |V \rangle_b ).
\end{align}
\end{subequations}

Before collected by a detector, the $\textbf{b}$ photon will only meet a PBS and be converted into $| H \rangle_{b} \Rightarrow | D_6 \rangle$ and $| V \rangle_{b} \Rightarrow | D_5 \rangle$. The $\mathbf{a}$ photon will meet a BS. Under the condition of only considering the evolution of the $\mathbf{a}$ photon, $| \phi^{\pm}\rangle$ will become $| \phi^{\pm}\rangle _ {cd}$ as follows
\begin{widetext}
\begin{subequations}
\begin{align}\label{03}
| \phi^{+}\rangle _ {cd}  = \frac{1}{4 \sqrt{2}} & [ |H \rangle_d |H \rangle_d +|V \rangle_d |V \rangle_d + |H \rangle_c |H \rangle_c + |V \rangle_c |V \rangle_c + 2 (|H \rangle_c |H \rangle_d + |V \rangle_c |V \rangle_d ) ], \\
| \phi^{-}\rangle _ {cd}  = \frac{1}{4 \sqrt{2}} & [ |H \rangle_d |H \rangle_d - |V \rangle_d |V \rangle_d + |H \rangle_c |H \rangle_c - |V \rangle_c |V \rangle_c + 2 (|H \rangle_c |H \rangle_d - |V \rangle_c |V \rangle_d )].
\end{align}
\end{subequations}
\end{widetext}

We can see from the above expressions that both $| \phi^{\pm}\rangle _ {cd}$ are still symmetric. If we let the two photons in the states $| \phi^{+}\rangle _ {cd}$ or $| \phi^{-}\rangle _ {cd}$ meet at a BS from its opposite sides, the photon bunching effect will occur again. We still would not be able to distinguish the states $| \phi^{+} \rangle$ and $| \phi^{-} \rangle$ from each other. However, if we let the $\mathbf{c}$ photon pass through a HWP, which can achieve the conversion as $| H \rangle_{c} \Rightarrow | V \rangle_{c \rm{H} }$ and $| V \rangle_{c} \Rightarrow | H \rangle_{c \rm{H} }$, and then $| \phi^{\pm}\rangle _ {cd} \Rightarrow | \phi^{\pm}\rangle _ {c {\rm H} d} $

\begin{widetext}
\begin{subequations}
\begin{align}\label{04}
| \phi^{+}\rangle _ {c {\rm H} d} = \frac{1}{4 \sqrt{2}} & [ |H \rangle_d |H \rangle_d +|V \rangle_d |V \rangle_d + |H \rangle_{c \rm{H}} |H \rangle_{c \rm{H}} + |V \rangle_{c \rm{H}} |V \rangle_{c \rm{H}} + 2 (|V \rangle_{c \rm{H}} |H \rangle_d + |H \rangle_{c \rm{H}} |V \rangle_d ) ], \\
| \phi^{-}\rangle _ {c {\rm H} d} = \frac{1}{4 \sqrt{2}} & [ |H \rangle_d |H \rangle_d - |V \rangle_d |V \rangle_d - |H \rangle_{c \rm{H}} |H \rangle_{c \rm{H}} + |V \rangle_{c \rm{H}} |V \rangle_{c \rm{H}} + 2 (|V \rangle_{c \rm{H}} |H \rangle_d - |H \rangle_{c \rm{H}} |V \rangle_d ) ].
\end{align}
\end{subequations}
\end{widetext}

Now $|\phi^{+}\rangle _ {c {\rm H} d}$ is still symmetric, whereas $|\phi^{-}\rangle _ {c {\rm H} d}$ has become antisymmetric. After meeting at a BS from its opposite sides, the photons in the state $|\phi^{+}\rangle _ {c {\rm H} d}$ end up in the same output port, the photons in the state $|\phi^{-}\rangle _ {c {\rm H} d}$ will exit in different output ports. Then, the states $|\phi^{+}\rangle$ and $|\phi^{-}\rangle$ can be distinguished from each other.

After passing through the whole symmetry-broken scheme, the four states will evolve into

\begin{widetext}
\begin{subequations}\label{05}
\begin{align}
| \psi ^{-}\rangle \Rightarrow \frac{1}{2 \sqrt{2}} & ( - |D_1 \rangle |D_5 \rangle -|D_2 \rangle |D_5 \rangle - |D_3 \rangle |D_5 \rangle +|D_4 \rangle |D_5 \rangle + |D_1 \rangle |D_6 \rangle + |D_2 \rangle |D_6 \rangle - |D_3 \rangle |D_6 \rangle +|D_4 \rangle |D_6 \rangle ), \\
| \psi ^{+}\rangle \Rightarrow \frac{1}{4 \sqrt{2}} & ( + |D_1 \rangle |D_1 \rangle + |D_2 \rangle |D_2 \rangle - |D_3 \rangle |D_3 \rangle - |D_4 \rangle |D_4 \rangle + 2 |D_1 \rangle |D_2 \rangle + 2 |D_3 \rangle |D_4 \rangle - 4 |D_5 \rangle |D_6 \rangle ), \\
| \phi ^{-}\rangle \Rightarrow \frac{1}{2 \sqrt{2}} & ( +|D_1 \rangle |D_3 \rangle -|D_1 \rangle  |D_4 \rangle + |D_2 \rangle |D_3 \rangle - |D_2 \rangle |D_4 \rangle + |D_5 \rangle |D_5 \rangle - |D_6 \rangle |D_6 \rangle ), \\
| \phi ^{+}\rangle \Rightarrow \frac{1}{4 \sqrt{2}} & ( + |D_1 \rangle |D_1 \rangle + |D_2 \rangle  |D_2 \rangle + |D_3 \rangle |D_3 \rangle +|D_4 \rangle |D_4 \rangle + 2 |D_1 \rangle  |D_2 \rangle - 2 |D_3 \rangle |D_4 \rangle - 2 |D_5 \rangle |D_5 \rangle - 2 |D_6 \rangle |D_6 \rangle ).
\end{align}
\end{subequations}
\end{widetext}

From the above expressions, we can see that when detector $D_5$ or $D_6$ fires in coincidence with any one of the other four detectors $D_1$$\sim$$D_4$, the input state can be identified as $| \psi ^{-}\rangle$. While $| \psi ^{+}\rangle$ can be assigned by coincidence between detectors $D_5$ and $D_6$. If one of $D_3$ and $D_4$ has a coincidence with any one of $D_1$ and $D_2$, the input state should be $| \phi ^{-}\rangle$. When $D_5$ and $D_6$ has no coincidence, at the same time, $D_1$ and $D_2$ or $D_3$ and $D_4$ has a coincidence, we can identify the input state as $| \phi ^{+} \rangle$.

With Eq.~(5), for one photon-pair, we can easily calculate the probability $P^{D_i}_{D_j}$ of the event that one photon reaches detector $D_i$ and the another one reaches detector $D_j$. The success probability $S_1$ for one photon-pair is the sum of the probabilities of the events, which can be used to distinguish one state from the others without error. For example, $S_1^{| \psi ^{-}\rangle} = P^{D_1}_{D_5} + P^{D_2}_{D_5} + P^{D_1}_{D_6} + P^{D_2}_{D_6} + P^{D_3}_{D_5} + P^{D_4}_{D_5} + P^{D_3}_{D_6} + P^{D_4}_{D_6} = 100\%$, $S_1^{| \psi ^{+}\rangle} = P^{D_5}_{D_6} = 50\%$, $S_1^{| \phi ^{-}\rangle} = P^{D_1}_{D_3} + P^{D_1}_{D_4} + P^{D_2}_{D_3} + P^{D_2}_{D_4} = 50\%$, and $S_1^{| \phi ^{+} \rangle} = 0$ calculated from Eqs.~(5a)--(5d), respectively. This means that it is impossible to realize the complete polarization Bell-state analysis for one photon pair only and the total probability of success is still 50\%, which is the same as the symmetric scheme with linear optics \cite{R16}. In the existence of $N$ photon-pairs, the success rate for the state $| X \rangle$ can be calculated as $S_N^{|X\rangle} = 1 - \left(1 - S_1^{|X\rangle} \right)^N$, where $P^{|X\rangle}_N = \left(1 - S_1^{|X\rangle} \right)^N$ stands for the probability that none of the $N$ photon pairs can be used to distinguish the state $| X \rangle$ from the others. So we can obtain $S_N^{| \psi ^{-}\rangle} = 1 - \left(1 - S_1^{| \psi ^{-}\rangle} \right)^ N =100\%$, $S_N^{| \psi ^{+}\rangle} = 1 - \left(1-S_1^{| \psi ^{+}\rangle} \right)^ N = 1 - 2^{-N}$, and $S_N^{| \phi ^{-}\rangle} = 1 - \left(1 - S_1^{| \phi ^{-}\rangle} \right)^ N = 1 - 2^{-N}$. Because two conditions are needed to identify the state $| \phi ^{+} \rangle$, the calculation of $S_N^{| \phi ^{+} \rangle}$ will be more complicated. Here we introduce an alternative method for calculating $S_N^{| \phi ^{+} \rangle}$, based on the complementary principle. After $S_N^{| \psi ^{-}\rangle}$, $S_N^{| \psi ^{+}\rangle}$, and $S_N^{| \phi ^{-}\rangle}$ among the four Bell states have been obtained, $S_N^{| \phi ^{+} \rangle}$ can be given by $S_N^{| \phi ^{+} \rangle} = 1 - \left(1 - S_N^{| \psi ^{-}\rangle} \right) - \left(1-S_N^{| \psi ^{+}\rangle} \right) - \left(1 - S_N^{| \phi ^{-}\rangle} \right) = 1 - 2^{-(N-1)}$.

\begin{figure}[hbpt]
\centering{\includegraphics[width=8.5cm]{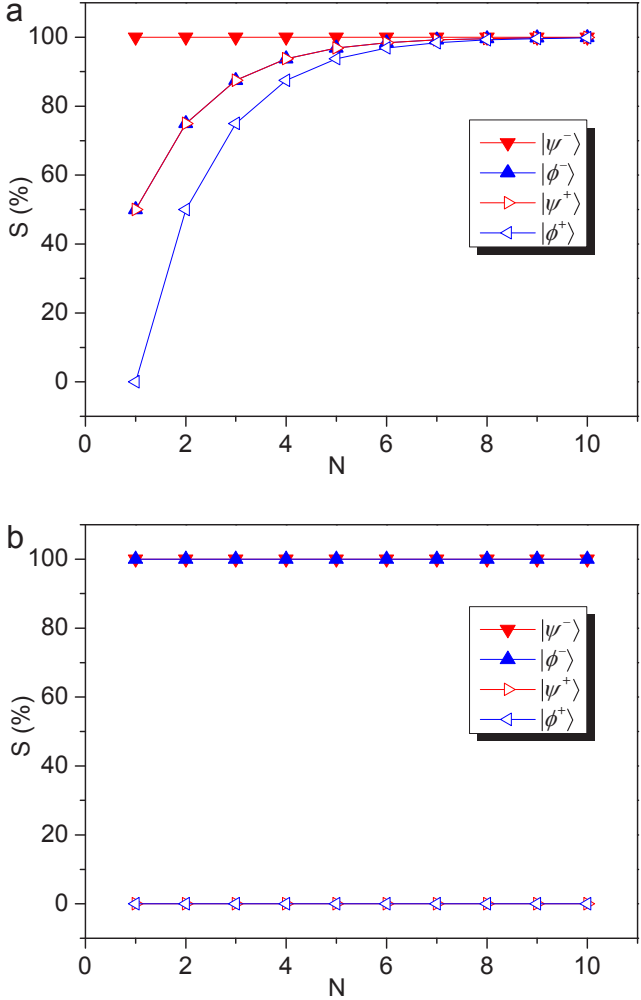}\\
  \caption{Impact of the number $N$ of photon pairs on the success probabilities $S_r$ for identifying the four Bell states $\mathbf{| \psi ^{-}\rangle}$, $| \psi ^{+}\rangle$, $| \phi ^{-}\rangle$, and $| \phi ^{+} \rangle$ in two schemes. \textbf{a} Symmetry-broken scheme. \textbf{b} Symmetric scheme.}}
\end{figure}

Figure~2a shows the dependence of the success rates for the states $| \psi ^{-}\rangle$, $| \psi ^{+}\rangle$, $| \phi ^{-}\rangle$, and $| \phi ^{+} \rangle$ on the number $N$ of photon pairs. Clearly, as the number of photon-pairs increases, the success probabilities for identifying the Bell-state analysis increase exponentially except for $S_r^{| \psi ^{-}\rangle} \equiv 100\% $ for the sate $| \psi ^{-}\rangle$. For instance, when $N = 4$ we have $S_N^{| \psi ^{-}\rangle} = 100\%$, $S_N^{| \psi ^{+}\rangle} = S_N^{| \phi ^{-}\rangle} = 93.8\%$, and $S_N^{| \phi ^{+} \rangle} = 87.5\%$. When $ N \geq 8 $, for the other three states, $S_r^{| \psi ^{+}\rangle} = S_r^{| \phi ^{-}\rangle}$ are higher than 99.6\% and $S_r^{| \phi ^{+} \rangle}$ is not lower than 99.2\%. This suggests that provided that $N \geq 8 $, all the four Bell states become completely distinguishable, while which cannot be realized by the symmetric scheme. This novel property originates from an essential difference of our symmetry-broken scheme from the symmetric scheme that our symmetry-broken scheme makes all the four Bell states be completely nondegenerate.

\noindent \textbf{Analysis.} We will use the method presented by Vaidman and Yoranin \cite{R29} to confirm the feasibility of our symmetry-broken scheme for distinguishing all the four polarization Bell states. The unitary linear evolution performed by our scheme for the four states $ |H \rangle_{a'}$, $ |H \rangle_{b'}$, $ |V \rangle_{a'}$, and $ |V \rangle_{b'}$ can be written in the following form
\begin{subequations}
\begin{align}\label{06}
|H \rangle_{a'} \Rightarrow  & \sum_{i} C^{1}_{i} |D_i \rangle, \\
|H \rangle_{b'} \Rightarrow  & \sum_{i} C^{2}_{i} |D_i \rangle,\\
|V \rangle_{a'} \Rightarrow  & \sum_{i} C^{3}_{i} |D_i \rangle, \\
|V \rangle_{b'} \Rightarrow  & \sum_{i} C^{4}_{i} |D_i \rangle.
\end{align}
\end{subequations}

\noindent Coefficients, $ C^{1}_{i}$, $ C^{2}_{i}$, $ C^{3}_{i}$, and $ C^{4}_{i}$, are easily calculated, as listed in Table~I. Given that the evolution of the photon in one state is independent of the photon in the other state, with Eq.~(6), the evolution of the Bell states can be written as follows
\begin{subequations}
\begin{align}\label{07}
| \psi ^{-}\rangle \Rightarrow & \sum_{ij} \alpha_{ij} |D_i \rangle |D_j \rangle, \\
| \psi ^{+}\rangle \Rightarrow & \sum_{ij} \beta_{ij} |D_i \rangle |D_j \rangle, \\
| \phi ^{-}\rangle \Rightarrow & \sum_{ij} \gamma_{ij} |D_i \rangle |D_j \rangle, \\
| \phi ^{+}\rangle \Rightarrow & \sum_{ij} \delta_{ij} |D_i \rangle |D_j \rangle.
\end{align}
\end{subequations}

\begin{table}
\caption{The values of the coefficients $ C^{1}_{i}$, $ C^{2}_{i}$, $ C^{3}_{i}$ and $ C^{4}_{i}$ for the symmetry-broken scheme shown in Fig. 1.}
\begin{tabular}{ccccccc}
\hline\hline
 & \quad $i=1$ & \quad $i=2$ & \quad $i=3$ & \quad $i=4$ & \quad $i=5$ & \quad $i=6$ \\
\hline
 $ C^{1}_{i} $ & \quad $ -\frac{1}{2\sqrt{2}} $ & \quad $ \frac{1}{2\sqrt{2}} $ & \quad $ \frac{1}{2\sqrt{2}} $ & \quad $- \frac{1}{2\sqrt{2}}$ & \quad  0   & $ \quad \frac{1}{\sqrt{2}} $ \\
 $ C^{2}_{i} $ & \quad $ \frac{1}{2\sqrt{2}} $ & \quad  $ \frac{1}{2\sqrt{2}} $  & \quad $ \frac{1}{2\sqrt{2}} $ & \quad $- \frac{1}{2\sqrt{2}} $ & \quad    0  & \quad $- \frac{1}{\sqrt{2}} $ \\
 $ C^{3}_{i} $ & \quad $ \frac{1}{2\sqrt{2}} $ & \quad $ \frac{1}{2\sqrt{2}} $ & \quad $- \frac{1}{2\sqrt{2}} $ & \quad $ \frac{1}{2\sqrt{2}} $ & \quad $ \frac{1}{\sqrt{2}} $  & \quad 0 \\
 $ C^{4}_{i} $ & \quad $ \frac{1}{2\sqrt{2}} $ & \quad $ \frac{1}{2\sqrt{2}} $ & \quad $- \frac{1}{2\sqrt{2}} $ & \quad $ \frac{1}{2\sqrt{2}} $ & \quad $- \frac{1}{\sqrt{2}} $ & \quad  0  \\
\hline
\hline
\end{tabular}
\end{table}

Assuming that only local detectors are used, only the product states can be detected. In the right hand side of Eq.~(7), the sum is only on pairs with $ i \leq j$. For $ i > j$, the state $ |D_i \rangle |D_j \rangle$ has been merged with $ |D_j \rangle |D_i \rangle$. For $ i = j $, we have
\begin{subequations}\label{08}
\begin{align}
\alpha_{ij} = \frac{1}{\sqrt{2}} \left( C^{1}_{i} C^{4}_{i} - C^{2}_{i} C^{3}_{i} \right), \\
\beta_{ij}  = \frac{1}{\sqrt{2}} \left( C^{1}_{i} C^{4}_{i} + C^{2}_{i} C^{3}_{i} \right), \\
\gamma_{ij} = \frac{1}{\sqrt{2}} \left( C^{1}_{i} C^{2}_{i} - C^{3}_{i} C^{4}_{i} \right), \\
\delta_{ij} = \frac{1}{\sqrt{2}} \left( C^{1}_{i} C^{2}_{i} + C^{3}_{i} C^{4}_{i} \right),
\end{align}
\noindent and for $ i \neq  j $, we obtain
\begin{align}
\alpha_{ij} = \frac{1}{\sqrt{2}} \left( C^{1}_{i} C^{4}_{j} + C^{1}_{j} C^{4}_{i} - C^{2}_{i} C^{3}_{j} - C^{2}_{j} C^{3}_{i} \right), \\
\beta_{ij}  = \frac{1}{\sqrt{2}} \left( C^{1}_{i} C^{4}_{j} + C^{1}_{j} C^{4}_{i} + C^{2}_{i} C^{3}_{j} + C^{2}_{j} C^{3}_{i} \right), \\
\gamma_{ij} = \frac{1}{\sqrt{2}} \left( C^{1}_{i} C^{2}_{j} + C^{1}_{j} C^{2}_{i} - C^{3}_{i} C^{4}_{j} - C^{3}_{j} C^{4}_{i} \right), \\
\delta_{ij} = \frac{1}{\sqrt{2}} \left( C^{1}_{i} C^{2}_{j} + C^{1}_{j} C^{2}_{i} + C^{3}_{i} C^{4}_{j} + C^{3}_{j} C^{4}_{i} \right).
\end{align}
\end{subequations}

With Table I and Eq.~(8), we can easily obtain the coefficients, $\alpha_{ij}$, $\beta_{ij}$, $\gamma_{ij}$, and $\delta_{ij}$. Substituting these coefficients into Eq.~(7), we will find the same results as Eq.~(5).

\noindent \textbf{Discussion.} If a symmetric scheme shown in Fig.~3 is used, the complete polarization Bell-state analysis is impossible as the confirmation below. Based on the transformation rules as mentioned above, the evolution of the four Bell states shows that $| \phi ^{+}\rangle$ and $| \phi ^{-}\rangle$ can be distinguished from each other, whereas $| \phi ^{+}\rangle$ from $| \psi ^{+}\rangle$ can never be distinguished. The detailed discussion is shown below.
\begin{figure}[bhp]
\centering{\includegraphics[width=8.5cm]{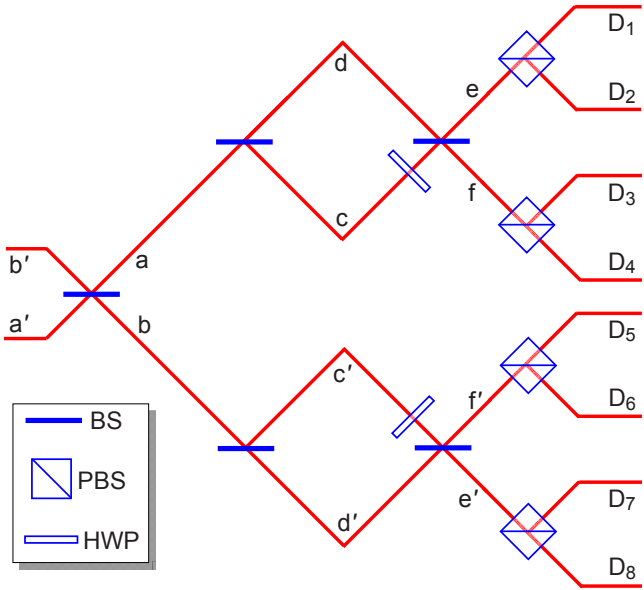}\\
  \caption{Symmetric scheme with linear optics only, for identifying the polarization Bell states. BS---50/50 beam splitter, PBS---polarization beam splitter, HWP---half-wave plate.}}
\end{figure}

Similar to the $\mathbf{a}$ photon described above, the $\mathbf{b}$ photon first meets a BS. After passing through a HWP, the $\mathbf{c'}$ photon undergoes the conversion as $| H \rangle_{c'} \Rightarrow | V \rangle_{c' \rm{H} }$ and $| V \rangle_{c'} \Rightarrow | H \rangle_{c' \rm{H} }$. Subsequently, the $\mathbf{c'}$ photon meets a BS again and interferes with the $\mathbf{d'}$ photon. At each of exits, the orthogonally polarized photons are separated by a PBS. After the photons pass through the whole symmetric scheme, the four states are evolved into
\begin{widetext}
\begin{subequations}\label{09}
\begin{align}
| \psi ^{-}\rangle \Rightarrow \frac{1}{2 \sqrt{2}} & ( - |D_1 \rangle |D_5 \rangle + |D_1 \rangle |D_6 \rangle - |D_2 \rangle |D_5 \rangle + |D_2 \rangle |D_6 \rangle + |D_3 \rangle |D_7 \rangle + |D_3 \rangle |D_8 \rangle - |D_4 \rangle |D_7 \rangle - |D_4 \rangle |D_8 \rangle ), \\
| \psi ^{+}\rangle \Rightarrow \frac{1}{2 \sqrt{2}} & ( + |D_1 \rangle |D_1 \rangle  + |D_2 \rangle |D_2 \rangle - |D_3 \rangle |D_3 \rangle - |D_4 \rangle |D_4 \rangle + |D_5 \rangle |D_5 \rangle + |D_6 \rangle |D_6 \rangle \nonumber \\
& - |D_7 \rangle |D_7 \rangle - |D_8 \rangle |D_8 \rangle + 2 |D_1 \rangle |D_2 \rangle + 2 |D_3 \rangle |D_4 \rangle - 2 |D_5 \rangle |D_6 \rangle - 2 |D_7 \rangle |D_8 \rangle ), \\
| \phi ^{-}\rangle \Rightarrow \frac{1}{2 \sqrt{2}} & ( +|D_1 \rangle |D_3 \rangle -|D_1 \rangle  |D_4 \rangle + |D_2 \rangle |D_3 \rangle -|D_2 \rangle |D_4 \rangle - |D_5 \rangle |D_7 \rangle - |D_5 \rangle |D_8 \rangle +|D_6 \rangle |D_7 \rangle + |D_6 \rangle |D_8 \rangle ), \\
| \phi ^{+}\rangle \Rightarrow \frac{1}{2 \sqrt{2}} & ( + |D_1 \rangle |D_1 \rangle  + |D_2 \rangle  |D_2 \rangle + |D_3 \rangle  |D_3 \rangle +|D_4 \rangle |D_4 \rangle - |D_5 \rangle |D_5 \rangle - |D_6 \rangle |D_6 \rangle \nonumber \\
& - |D_7 \rangle |D_7 \rangle - |D_8 \rangle |D_8 \rangle+ 2 |D_1 \rangle  |D_2 \rangle - 2 |D_3 \rangle |D_4 \rangle + 2 |D_5 \rangle |D_6 \rangle - 2 |D_7 \rangle |D_8 \rangle ).
\end{align}
\end{subequations}
\end{widetext}

The above expressions show that the state $| \psi ^{-}\rangle $ can be identified when one of the detectors $|D_1 \rangle$ and $|D_2 \rangle $ fires in coincidence with any one of the detectors $|D_5 \rangle$ and $|D_6 \rangle $, or one of $|D_3 \rangle$ and $|D_4 \rangle $ fires in coincidence with any one of $|D_7 \rangle$ and $|D_8 \rangle $. The state $| \phi ^{-}\rangle $ can be designated when one of $|D_1 \rangle$ and $|D_2 \rangle$ has a coincidence with any one of $|D_3 \rangle$ and $|D_4 \rangle$, or one of $|D_5 \rangle$ and $|D_6 \rangle$ has a coincidence with any one of  $|D_7 \rangle$ and $|D_8 \rangle$. However, the states $| \psi ^{+}\rangle $ and $| \phi ^{+}\rangle $ have always the degenerate signal, implying that the two states cannot be identified from each other.

Using the same method mentioned above, we easily calculate with Eqs.~(9a)--(9d) the success probabilitys for the four Bell states. For example, we have $S_1^{| \psi ^{-}\rangle} = P^{D_1}_{D_5} + P^{D_2}_{D_5} + P^{D_1}_{D_6} + P^{D_2}_{D_5} + P^{D_3}_{D_7} + P^{D_4}_{D_7} + P^{D_3}_{D_8} + P^{D_4}_{D_8} = 100\%$, $S_1^{| \psi ^{+}\rangle} = 0$, $S_1^{| \phi ^{-}\rangle} = P^{D_1}_{D_3} + P^{D_1}_{D_4} + P^{D_2}_{D_3} + P^{D_2}_{D_4} + P^{D_5}_{D_7} + P^{D_5}_{D_8} + P^{D_6}_{D_7} + P^{D_6}_{D_8} = 100\%$, and $S_1^{| \phi ^{+} \rangle} = 0$ for the case of one photon pair, respectively. What's more, $S_N^{| \psi ^{+}\rangle} = S_1^{| \psi ^{+}\rangle}$ =100\%, $S_N^{| \psi ^{+}\rangle} =S_1^{| \psi ^{+}\rangle} = 0$, $S_N^{| \phi ^{-}\rangle} =S_1^{| \phi ^{-}\rangle} =100\%$, and $S_N^{| \phi ^{+} \rangle} = S_1^{| \phi ^{+} \rangle} = 0$ for the case of $N$ photon pairs. As shown in Fig.~2b, the success probabilitys for four Bell states are independent of the number $N$ of photon pairs. Consequently, no matter how many entangled photon-pairs, the complete polarization Bell-state analysis can never be accomplished by the symmetric scheme. As a result, only three distinct classes are distinguishable for the polarization Bell states by the symmetric scheme with linear optics, implying that it is possible to encode two photons into $\log_2 3 = 1.585$ bits of postselection channel capacity.

The method presented in Ref.~29 can still be used to confirm the fact that the symmetric scheme composed of linear optical elements cannot distinguish the four Bell states completely. For the symmetric scheme, Eqs.~(6)-(8) are still valid. As listed in Table II, the unique difference is that the values of the coefficients $ C^{1}_{i} $, $ C^{2}_{i} $, $ C^{3}_{i} $, and $ C^{4}_{i} $  have been changed. With Table II, we easily calculate the values of $\alpha_{ij}$, $\beta_{ij}$, $\gamma_{ij}$, and $ \delta_{ij} $. Substituting these coefficients into Eq.~(7), we will find the same results as Eq.~(9).

\begin{table*}
\caption{The values of the coefficients, $ C^{1}_{i}$, $ C^{2}_{i}$, $ C^{3}_{i}$ and $ C^{4}_{i}$, for the symmetric scheme shown in Fig.~3.}
\begin{tabular}{ccccccccc}
\hline\hline
 & \quad $i=1$ & \quad $i=2$ & \quad $i=3$ & \quad $i=4$ & \quad $i=5$ & \quad $i=6$ & \quad $i=7$ & \quad $i=8$ \\
\hline
$ C^{1}_{i} $ & \quad $ \frac{1}{2\sqrt{2}} $ & \quad $ \frac{1}{2\sqrt{2}} $ & \quad $ \frac{1}{2\sqrt{2}} $ & \quad $- \frac{1}{2\sqrt{2}} $ & \quad $ - \frac{1}{2\sqrt{2}} $ \quad & $ \frac{1}{2\sqrt{2}} $ & \quad $- \frac{1}{2\sqrt{2}} $ & \quad $ - \frac{1}{2\sqrt{2}} $ \\
$ C^{2}_{i} $ & \quad $ \frac{1}{2\sqrt{2}} $ & \quad $ \frac{1}{2\sqrt{2}} $ & \quad $ \frac{1}{2\sqrt{2}} $ & \quad $- \frac{1}{2\sqrt{2}} $ & \quad $ \frac{1}{2\sqrt{2}} $ & \quad $- \frac{1}{2\sqrt{2}} $ & \quad $ \frac{1}{2\sqrt{2}} $ & \quad $ \frac{1}{2\sqrt{2}} $ \\
$ C^{3}_{i} $ & \quad $ \frac{1}{2\sqrt{2}} $ & \quad $ \frac{1}{2\sqrt{2}} $ & \quad $- \frac{1}{2\sqrt{2}} $ & \quad $ \frac{1}{2\sqrt{2}} $ & \quad $ \frac{1}{2\sqrt{2}} $ & \quad $ - \frac{1}{2\sqrt{2}} $ & \quad $ - \frac{1}{2\sqrt{2}} $ & \quad $- \frac{1}{2\sqrt{2}} $ \\
$ C^{4}_{i} $ & \quad $ \frac{1}{2\sqrt{2}} $ & \quad $ \frac{1}{2\sqrt{2}} $ & \quad $- \frac{1}{2\sqrt{2}} $ & \quad $ \frac{1}{2\sqrt{2}} $ \quad & $- \frac{1}{2\sqrt{2}} $ & \quad $ \frac{1}{2\sqrt{2}} $ & \quad $ \frac{1}{2\sqrt{2}} $ & \quad $ \frac{1}{2\sqrt{2}} $ \\
\hline\hline
\end{tabular}
\end{table*}

In summary, we have proposed a symmetry-broken scheme for the polarization Bell-state analysis with linear optics only, and no auxiliary photons, feedback technique and nonlinear interaction are required. The fundamental principle is based on the Hong-Ou-Mandel interference for distinguishing the symmetric and antisymmetric states. Although it has been proved that a complete deterministic Bell-state measurement for every entangled photon-pair is impossible by using the linear optics only \cite{R29,R30}, the symmetry-broken scheme we presented here can break completely the degeneracy of the Bell states. This novel feature results in the fact that as the photon-pair number $N$ increases, the success probabilities for the Bell-state analysis will increase exponentially. Under the condition that the entangled photon-pair number $N \geq 8$, the success probabilities of identifying all the four Bell states are larger than 99.2\%, which implying that the four Bell states are completely discriminated from each other and that encoding into two photons near $\log_2 4 = 2$ bits of postselection channel capacity is possible. Our symmetry-broken scheme with linear optics only should be a crucial breakthrough in the entangled Bell-state analysis. The symmetry breaking plays an indispensable role in our scheme as in other physical systems. Our scheme is simpler and feasible with respect to the current technology for the realization of the near-complete polarization Bell state analysis. In addition, our scheme also provides us with an alternative route of separating the entangled photons and reentangling the photons. Our scheme should be of great significance for practical applications in quantum communication protocols and so on.

\noindent \textbf{Acknowledgements}

\noindent This work is supported by National Basic Research Program (973 Program) of China under Grant No. 2012CB921900, National Natural Science Foundation of China under Grant Nos. 11534006, 11274183 and 11374166, National scientific instrument and equipment development project 2012YQ17004, and Tianjin research program of application foundation and advanced technology 13JCZDJC33800 and 12JCYBJC10700.

\vspace{0.5cm}
\noindent \textbf{Appendix: Methods}

\noindent In both the symmetry-broken and symmetric schemes, the past axis for any HWP is oriented along the angular bisector between the horizontal and vertical directions. Such HWP can convert the horizontal polarization into the vertical one and vice versa. For any PBS, the horizontally polarized photons are transmitted, whereas the vertically polarized ones are reflected. The evolution of the photon states can be traced, based on the Hong-Ou-Mandel interference principle \cite{R33} and the theoretical method presented by Vaidman and Yoran \cite{R29}.

\noindent \textbf{The evolution of the photon states in the symmetry-broken scheme.} When the $\mathbf{a'}$ and $\mathbf{b'}$ photons meet a BS from its opposite sides, after undergoing the unitary $U(2)$ operation, the states of the horizontally polarized $\mathbf{a'}$ and $\mathbf{b'}$ photons are transferred into $|H \rangle_{a'} \Rightarrow \frac{1}{\sqrt{2}} (|H \rangle_{a} + |H \rangle_{b})$ and $|H \rangle_{b'} \Rightarrow \frac{1}{\sqrt{2}} (|H \rangle_{a} - |H \rangle_{b})$, respectively.

When the horizontally polarized $\mathbf{a}$ photon meets another BS, its state evolves into $|H \rangle_{a} \Rightarrow \frac{1}{\sqrt{2}} (|H \rangle_{c} + |H \rangle_{d})$. The horizontally polarized $\mathbf{c}$ photon is converted into the vertically polarized one by HWP, as $|H \rangle_{c} \Rightarrow |V \rangle_{c}$. When the horizontally polarized $\mathbf{d}$ and vertically polarized $\mathbf{c}$ photons meet at another BS from its opposite sides, their states are converted into $|H \rangle_{d} \Rightarrow \frac{1}{\sqrt{2}} (|H \rangle_{e} - |H \rangle_{f})$ and $|V \rangle_{c} \Rightarrow \frac{1}{\sqrt{2}} (|V \rangle_{e} + |V \rangle_{f})$, respectively, after undergoing the unitary $U(2)$ operation. After the $\mathbf{e}$ ($\mathbf{f}$) photon passes through a PBS, the horizontally and vertically polarized $\mathbf{e}$ ($\mathbf{f}$) photons are transmitted and reflected, and then arrive at detectors $D_1$ and $D_2$ ($D_4$ and $D_3$), respectively, i.e., $|H \rangle_{e} \Rightarrow |D_1 \rangle$ and $|V \rangle_{e} \Rightarrow |D_2 \rangle$ ($|H \rangle_{f} \Rightarrow |D_4 \rangle$ and $|V \rangle_{f} \Rightarrow |D_3 \rangle$). After the horizontally polarized $\mathbf{b}$ photon passes through a PBS, it will be transmitted and will arrives at the detector $D_6$, i.e., $|H \rangle_{b} \Rightarrow |D_6 \rangle$. As mentioned above, after the evolution in the whole symmetry-broken scheme, the states of the horizontally polarized $\mathbf{a'}$ and $\mathbf{b'}$ photons are evolved into
\noindent
\begin{subequations}\label{10}
\begin{align}
|H \rangle_{a'} \Rightarrow & \frac{1}{2 \sqrt{2}} (|H \rangle_{e} + |V \rangle_{e} + |V \rangle_{f} - |H \rangle_{f}) + \frac{1}{\sqrt{2}} |H \rangle_{b}, \\
|H \rangle_{b'} \Rightarrow & \frac{1}{2 \sqrt{2}} (|H \rangle_{e} + |V \rangle_{e} + |V \rangle_{f} - |H \rangle_{f}) - \frac{1}{\sqrt{2}} |H \rangle_{b},
\end{align}

Referencing the cases of  horizontally polarized $\mathbf{a'}$ and $\mathbf{b'}$ photons, the state evolutions of the vertically polarized $\mathbf{a'}$ and $\mathbf{b'}$ photons in the whole symmetry-broken scheme are easily given provided $|H \rangle$ and $|V \rangle$ in Eqs. (10a) and (10b) are exchanged with each other (i.e., $|H \rangle \Rightarrow |V \rangle$ and $|V \rangle \Rightarrow |H \rangle$), as follows

\noindent
\begin{align}
|V \rangle_{a'} \Rightarrow & \frac{1}{2 \sqrt{2}} (|V \rangle_{e} + |H \rangle_{e} + |H \rangle_{f} - |V \rangle_{f}) + \frac{1}{\sqrt{2}} |V \rangle_{b}, \\
|V \rangle_{b'} \Rightarrow & \frac{1}{2 \sqrt{2}} (|V \rangle_{e} + |H \rangle_{e} + |H \rangle_{f} - |V \rangle_{f}) - \frac{1}{\sqrt{2}} |V \rangle_{b}.
\end{align}
\end{subequations}

Using specific correspondences, $|H \rangle_{e} \Rightarrow |D_1 \rangle$, $|V \rangle_{e} \Rightarrow |D_2 \rangle$, $|V \rangle_{f} \Rightarrow |D_3 \rangle$, $|H \rangle_{f} \Rightarrow |D_4 \rangle$, $|V \rangle_{b} \Rightarrow |D_5 \rangle$ and $|H \rangle_{b} \Rightarrow |D_6 \rangle$, the expression (10) can be rewritten as follows
\noindent
\begin{subequations}\label{11}
\begin{align}
|H \rangle_{a'} \Rightarrow & \frac{1}{2 \sqrt{2}} (|D_1 \rangle + |D_2 \rangle + |D_3 \rangle - |D_4 \rangle) + \frac{1}{\sqrt{2}} |D_6 \rangle, \\
|H \rangle_{b'} \Rightarrow & \frac{1}{2 \sqrt{2}} (|D_1 \rangle + |D_2 \rangle + |D_3 \rangle - |D_4 \rangle) - \frac{1}{\sqrt{2}} |D_6 \rangle, \\
|V \rangle_{a'} \Rightarrow & \frac{1}{2 \sqrt{2}} (|D_1 \rangle + |D_2 \rangle - |D_3 \rangle + |D_4 \rangle) + \frac{1}{\sqrt{2}} |D_5 \rangle, \\
|V \rangle_{b'} \Rightarrow & \frac{1}{2 \sqrt{2}} (|D_1 \rangle + |D_2 \rangle - |D_3 \rangle + |D_4 \rangle) - \frac{1}{\sqrt{2}} |D_5 \rangle.
\end{align}
\end{subequations}

\noindent Comparing the expression (11) with the expression (6), the coefficients $ C^{1}_{i}$, $ C^{2}_{i}$, $ C^{3}_{i}$, and $ C^{4}_{i}$ in the symmetry-broken scheme can be easily obtained, as listed in Table I.

\noindent \textbf{The evolution of the photon states in the symmetric scheme.} In this case, the unique difference from the symmetry-broken scheme is the different state evolution of the $\mathbf{b}$ photon. In the symmetry-broken scheme, the horizontally and vertically polarized $\mathbf{b}$ photons arrive at $|D_6 \rangle$ and $|D_5 \rangle$ ($|H \rangle_{b} \Rightarrow |D_6 \rangle$ and $|V \rangle_{b} \Rightarrow |D_5 \rangle$), respectively, after passing through a PBS.

In the symmetric scheme shown in Fig. 3, however, the $\mathbf{b}$ photon undergoes the state evolution as follows. After passing through a BS, the horizontally polarized $\mathbf{b}$ photon evolves into $|H \rangle_{b} \Rightarrow \frac{1}{\sqrt{2}} (|H \rangle_{c'} - |H \rangle_{d'})$. The horizontally polarized $\mathbf{c'}$ photon is converted into the vertically polarized one by HWP, as $|H \rangle_{c'} \Rightarrow |V \rangle_{c'}$. The horizontally polarized $\mathbf{d'}$ and vertically polarized $\mathbf{c'}$ photons meets another BS from its opposite sides, hence, their states are converted into $|H \rangle_{d'} \Rightarrow \frac{1}{\sqrt{2}} (|H \rangle_{e'} + |H \rangle_{f'})$ and $|V \rangle_{c'} \Rightarrow \frac{1}{\sqrt{2}} (- |V \rangle_{e'} + |V \rangle_{f'})$, respectively. When the $\mathbf{e'}$ ($\mathbf{f'}$) photon passes through a PBS, the horizontally and vertically polarized $\mathbf{e'}$ ($\mathbf{f'}$) photons are transmitted and reflected, and then arrive at detectors $D_8$ and $D_7$ ($D_5$ and $D_6$), respectively. Finally, the horizontally polarized $\mathbf{b}$ photon evolves into
\noindent
\begin{subequations}\label{12}
\begin{align}
|H \rangle_{b} \Rightarrow \frac{1}{2} (- |H \rangle_{e'} - |V \rangle_{e'} - |H \rangle_{f'} + |V \rangle_{f'}).
\end{align}

\noindent For the vertically polarized $\mathbf{b}$ photon, the state evolution is easily given by exchanging $|H \rangle \Rightarrow |V \rangle$ and $|V \rangle \Rightarrow |H \rangle$ in the above expression of $|H \rangle_{b}$ as follows
\noindent
\begin{align}
|V \rangle_{b} \Rightarrow \frac{1}{2} (- |V \rangle_{e'} - |H \rangle_{e'} - |V \rangle_{f'} + |H \rangle_{f'}).
\end{align}
\end{subequations}

Substituting $|H \rangle_{b}$ and $|V \rangle_{b}$ in Eq. (12) into Eq. (10), and using the special correspondences $|H \rangle_{f'} \Rightarrow |D_5 \rangle$, $|V \rangle_{f'} \Rightarrow |D_6 \rangle$, $|V \rangle_{e'} \Rightarrow |D_7 \rangle$ and $|H \rangle_{e'} \Rightarrow |D_8 \rangle$, the state evolutions of $\mathbf{a'}$ and $\mathbf{b'}$ photons are written as follows
\noindent
\begin{widetext}
\begin{subequations}\label{13}
\begin{align}
|H \rangle_{a'} \Rightarrow & \frac{1}{2 \sqrt{2}} (|D_1 \rangle + |D_2 \rangle + |D_3 \rangle - |D_4 \rangle - |D_5 \rangle + |D_6 \rangle - |D_7 \rangle - |D_8 \rangle), \\
|H \rangle_{b'} \Rightarrow & \frac{1}{2 \sqrt{2}} (|D_1 \rangle + |D_2 \rangle + |D_3 \rangle - |D_4 \rangle + |D_5 \rangle - |D_6 \rangle + |D_7 \rangle + |D_8 \rangle, \\
|V \rangle_{a'} \Rightarrow & \frac{1}{2 \sqrt{2}} (|D_1 \rangle + |D_2 \rangle - |D_3 \rangle + |D_4 \rangle + |D_5 \rangle - |D_6 \rangle - |D_7 \rangle - |D_8 \rangle), \\
|V \rangle_{b'} \Rightarrow & \frac{1}{2 \sqrt{2}} (|D_1 \rangle + |D_2 \rangle - |D_3 \rangle + |D_4 \rangle - |D_5 \rangle + |D_6 \rangle + |D_7 \rangle + |D_8 \rangle).
\end{align}
\end{subequations}
\end{widetext}

\noindent Comparing the expression (13) with the expression (6), the coefficients $ C^{1}_{i}$, $ C^{2}_{i}$, $ C^{3}_{i}$ and $ C^{4}_{i}$ in the symmetric scheme can be easily obtained, as listed in Table II.

\end{document}